\documentclass[aps,groupedaddress,showpacs,floatfix]{revtex4}
\usepackage{amssymb}
\usepackage{amsmath}
\usepackage{graphicx}
\DeclareMathOperator{\sgn}{sgn}
\DeclareMathOperator{\Ai}{Ai}
\begin{document}

\title{Distribution function of the endpoint
fluctuations of one-dimensional directed polymers
in a random potential}

\author{Victor Dotsenko}

\affiliation{LPTMC, Universit\'e Paris VI, 75252 Paris, France}

\affiliation{L.D.\ Landau Institute for Theoretical Physics,
 119334 Moscow, Russia}

\date{\today}

\begin{abstract}
The explicit expression for the the probability
distribution function of the endpoint fluctuations of one-dimensional directed polymers
in random potential is derived  in terms of the Bethe ansatz replica
technique by mapping the replicated problem to the $N$-particle quantum boson
system with attractive interactions.

\end{abstract}


\maketitle

\medskip

\section{Introduction}

One-dimensional directed polymers in a quenched random potential
and equivalent problem of the solutions of the
KPZ-equation \cite{KPZ} describing the growth in time of an interface
in the presence of noise
have been the subject of intense investigations during the past two
decades (see e.g. \cite{hh_zhang_95,burgers_74,kardar_book,hhf_85,numer1,numer2,kardar_87,bouchaud-orland,Brunet-Derrida,Johansson,Prahofer-Spohn,Ferrari-Spohn1,Corwin}).
The model of directed polymers describes
an elastic string  directed along the $\tau$-axis
within an interval $[0,t]$. Randomness enters the problem
through a disorder potential $V[\phi(\tau),\tau]$, which competes against
the elastic energy.  The system is defined by the Hamiltonian
\begin{equation}
   \label{1}
   H[\phi(\tau), V] = \int_{0}^{t} d\tau
   \Bigl\{\frac{1}{2} \bigl[\partial_\tau \phi(\tau)\bigr]^2
   + V[\phi(\tau),\tau]\Bigr\};
\end{equation}
where the disorder potential $V[\phi,\tau]$
is Gaussian distributed with a zero mean $\overline{V(\phi,\tau)}=0$
and the $\delta$-correlations:
\begin{equation}
   \label{2}
{\overline{V(\phi,\tau)V(\phi',\tau')}} = u \delta(\tau-\tau') \delta(\phi-\phi')
\end{equation}
Here the parameter $u$ describes the strength of the disorder.

In what follows we consider the problem in which the polymer is fixed
at the origin, $\phi(0)=0$ and it is free at $\tau = t$.
In other words, for a given realization of the random potential
$V$ the partition function of the considered system is:
\begin{equation}
\label{3}
   Z = \int_{-\infty}^{+\infty} dx \; Z(x) \; = \; \exp\{-\beta F\}
\end{equation}
where
\begin{equation}
\label{4}
   Z(x) = \int_{\phi(0)=0}^{\phi(t)=x}
              {\cal D} \phi(\tau)  \;  \mbox{\Large e}^{-\beta H[\phi]}
\end{equation}
is the partition function of the system with the fixed boundary condition,
$\phi(t)=x$ and $F$ is the total free energy.
Besides the usual extensive   part $f_{0} t$
(where $f_{0}$ is the linear free energy density),
the total free energy $F$ of such system is known to contain the disorder dependent
fluctuating contribution which in the limit of large $t$
scales  as
$t^{1/3}$ (see e.g. \cite{hhf_85,numer1,numer2,kardar_87}).
In other words, in the limit of large $t$ the total (random) free energy of the system
can be represented as $F =  f_{0} t  +  c  t^{1/3} f$,
where $c$ is a non-universal parameter, which depends on the temperature and the strength of disorder,
and $f$ is the random quantity which in the thermodynamic
limit $t\to\infty$ is described by a non-trivial universal
distribution function $P(f)$. The trivial self-averaging  contribution $f_{0}t$ to the free energy can be eliminated from the further study by the simple redefinition of the partition function,
$Z = \exp\{-\beta f_{0} t\} \, \tilde{Z}$, so that $\tilde{Z}  =  \exp\{-\lambda f\}$, 
where $\lambda =  \beta \, c \, t^{1/3}$.
Thus, to simplify notations the contribution $f_{0}t$
will be just dropped out in the further calculations.

For the problem with the zero boundary conditions, $\phi(0)=\phi(t)=0$,
the  distribution function $P(f)$ was demonstrated to be described by
the Gaussian Unitary Ensemble (GUE) Tracy-Widom distribution
\cite{KPZ-TW1,KPZ-TW2,BA-replicas,LeDoussal1}.
On the other hand, the free energy
distribution function of the directed polymers with
the free boundary conditions, eqs.(\ref{1})-(\ref{4}),
was shown to be given by the Gaussian Orthogonal Ensemble (GOE)
Tracy-Widom distribution \cite{LeDoussal2,goe}
In the course  of these derivations
rather efficient Bethe ansatz replica technique has been developed
\cite{BA-replicas,LeDoussal1,LeDoussal2,goe}.
Here in terms of this technique we are going to study on the statistical
properties of the transverse fluctuations of the directed polymers.
The scaling properties of the typical value of the endpoint  deviations,
$\phi(t)$, at large times is well known:
$ \overline{\langle\phi(t)\rangle^{2}}
\; \propto \; t^{4/3} \; \; $
(here $\langle ...\rangle$ denotes the thermal average
and $\overline{(...)}$ is the average over the disorder potential,
eq.(\ref{2})) \cite{hhf_85,numer1,numer2,kardar_87}.
Much more interesting object is the probability distribution function
$P(x)$ for the rescaled quantity $x = \phi(t)/t^{2/3}$
which is expected to become a universal function in the limit $t \to \infty$.
Recently this function has been derived in terms of the so called
maximal point of the $\mbox{Airy}_{2}$ process minus a parabola
\cite{math1, math2, math3}, which is believed to descibe the scaling limit
of the endpoint of the directed polymers in a random potential.
The long-standing conjecture that the top line of the Airy line ensemble minus 
a parabola attains its maximum at a unique point was recently proved in \cite{Corwin-Hammond}.
The obtained explicit expression for $P(x)$ turned out to be
rather complicated and its analytic properties is not so easy
to analyze although the asymptotic behavior of this function
is already known: $P(x\to\infty) \sim  \exp\{- |x|^{3}/12\}$\cite{math2}.

In this work the explicit form of the
distribution function of the directed polymer's endpoint fluctuations
will be derived in terms of the Bethe ansatz replica technique.
The distribution function we are going to consider is defined
as follows:
\begin{equation}
\label{9}
W(x) \; = \; \lim_{t\to\infty} \;
\mbox{Prob}\bigl[\phi(t)t^{-2/3} \; > \; x\bigr] \; = \;
\int_{x}^{\infty} \; dx' \; P(x')
\end{equation}
This function gives the probability that the rescaled value of the
polymer's right endpoint $\phi(t)/t^{2/3}$ is  bigger
than a given value $x$.
In this paper it will be shown that (see eqs.(\ref{82})-(\ref{87}) below)
\begin{equation}
 \label{10}
W(x) \; = \; \int_{-\infty}^{+\infty} df \;
F_{1}(-f)
\int_{0}^{+\infty} d\omega \int_{0}^{+\infty} d\omega'
\bigl(\hat{1} - \hat{B}_{-f}\bigr)^{-1}(\omega,\omega') \;
\Phi(\omega',\omega; \; f,x)
\end{equation}
Here $\hat{B}_{-f}$ is the integral operator with the kernel
$B_{-f}(\omega,\omega') = \Ai(\omega+\omega'-f) \; \;  (\omega, \omega' \, > \, 0)$,
the function $F_{1}(-f) = \det\bigl[\hat{1} -  \hat{B}_{-f} \bigr]$
is the GOE Tracy-Widom distribution and  $\bigl(\hat{1} - \hat{B}_{-f}\bigr)^{-1}(\omega,\omega')$
denotes the kernel of the inverse operator $\bigl(\hat{1} - \hat{B}_{-f}\bigr)^{-1}$
in $\omega$ and $\omega'$ (note that since $F_{1}(-f) > 0$ for all real $f$, the operator
$\bigl(\hat{1} - \hat{B}_{-f}\bigr)$ is invertible). The function $\Phi(\omega',\omega; \; f,x)$
is defined as follows:

\begin{eqnarray}
 \nonumber
\Phi(\omega,\omega'; \; f,x) = -\frac{1}{2} \int_{0}^{+\infty} dy \;
&\Biggl[&
\Bigl(
\frac{\partial}{\partial \omega} - \frac{\partial}{\partial \omega'}
\Bigr)
\Psi\bigl(\omega - \frac{1}{2}f + y ; \; x \bigr)
\Psi\bigl(\omega' - \frac{1}{2}f + y ; \; -x\bigr) +
\\
\nonumber
\\
&+&
\Bigl(
\frac{\partial}{\partial \omega} + \frac{\partial}{\partial \omega'}
\Bigr)
\Psi\bigl(\omega - \frac{1}{2}f - y ; \; x\bigr)
\Psi\bigl(\omega' - \frac{1}{2}f + y ; \; -x\bigr)
\Biggr]
\label{11}
\end{eqnarray}
where
\begin{equation}
\label{11a}
\Psi(\omega; x) \; = \;
2^{1/3} \mbox{Ai}\Bigl[2^{1/3}\bigl(\omega + \frac{1}{8} x^{2}\bigr)\Bigr] \, 
\exp\bigl\{-\frac{1}{2} \omega x\bigr\}
\end{equation}
The above result looks quite similar to the one obtained in \cite{math1},
although  at the moment I am not able to provide the proof that
these results are indeed 
{\it the same}$^{1}$\footnotetext[1]{After the paper has been accepted for publication 
I have learned that the equivalence of the result of this work, eqs.(\ref{10})-(\ref{11a}),
and the one of Refs.\cite{math1, math2, math3} has been proved in the recent paper
\cite{Bothner-Liechty}}.
In any case, the above expressions, eqs.(\ref{10})-(\ref{11a}), 
for the probability function $P(x)$ look as complicated 
as the ones obtained in Refs.\cite{math1, math2, math3}, 
and for the moment its analytic properties are not clear.

The paper is organized as follows.
   In Section II we define the distribution function $W(x)$ via the two-point
free energy distribution function $V_{x}(f_{1},f_{2})$ which give the probability
that the free energy of the polymer with the endpoint located above a position $x$
is bigger than a given value $f_{1}$, while the free energy of the polymer with
the endpoint located below the position $x$
is bigger than a given value $f_{2}$.
   In Section III the function $V_{x}(f_{1},f_{2})$ is defined by
mapping the considered problem to the
one-dimensional $N$-particle system of quantum bosons with attractive
$\delta$-interactions.
   In Section IV the explicit expression for the probability function
$V_{x}(f_{1},f_{2})$ is obtained in terms of the Bethe ansatz replica technique.
   Finally, in Section V the result eqs.(\ref{10})-(\ref{11a}) is derived.
Conclusions and future perspectives are discussed in Section VI.

\section{The endpoint probability distribution function}

In terms of the partition function $Z(x)$, eqs.(\ref{4}),
the probability distribution function of the polymer's endpoint $W(x)$,
eq.(\ref{9}), can be defined as follows:
\begin{eqnarray}
\nonumber
W(x) & = &
\lim_{t\to\infty} \;
\overline{\Biggl(
\int_{x}^{+\infty} dx'
\frac{Z(x')}{\int_{-\infty}^{+\infty} dx' \; Z(x')}
\Biggr)}
\\
\nonumber
\\
& = &
\lim_{t\to\infty} \;
\overline{\Biggl(
\frac{Z^{(+)}(x)}{Z^{(+)}(x) + Z^{(-)}(x)}
\Biggr)}
\label{12}
\end{eqnarray}
where
\begin{eqnarray}
\label{13}
Z^{(-)}(x) & \equiv &
\int_{-\infty}^{x} dx' \; Z(x')
\; = \;
\exp\{-\lambda f_{(-)}\}
\\
\nonumber
\\
Z^{(+)}(x) & \equiv &
\int_{x}^{+\infty} dx' \; Z(x')
\; = \;
\exp\{-\lambda f_{(+)}\}
\label{14}
\end{eqnarray}
where the parameter $\lambda\propto t^{1/3}$  and
$f_{(\pm)}$ are the free energies of the polymers with the
endpoint $\phi(t)$ located correspondingly above and below  a given
position $x$.
According to these definitions we find
\begin{equation}
\label{15}
W(x) \; = \;
\lim_{\lambda\to\infty} \;
\frac{\exp\{-\lambda f_{(+)}\} }{
\exp\{-\lambda f_{(-)}\} \,+ \, \exp\{-\lambda f_{(+)}\} }
\; =\;  \left\{
\begin{array}{ll}
0 , \; \; \; \;  \mbox{for} \; \;  f_{(-)} < f_{(+)} \\
\\
1 , \; \; \; \;  \mbox{for} \; \;  f_{(-)} > f_{(+)}
\end{array}
\right.
\end{equation}
Let us introduce the joint probability density function
${\cal P}_{x}\bigl[f_{(+)} ; f_{(-)}\bigr]$.
By definition the quantity
${\cal P}_{x}\bigl[f_{(+)} ; f_{(-)}\bigr] \, df_{(+)} \, df_{(-)}$
gives the probability that the free energy of the polymer
with the endpoint located below $x$ is equal to $f_{(-)}$
(within the interval $df_{(-)}$), while
the free energy of the polymer
with the endpoint located above $x$ is equal to $f_{(+)}$
(within the interval $df_{(+)}$).
Thus, according to eq.(\ref{15}),
\begin{equation}
\label{16}
W(x) \; = \;
\int_{-\infty}^{+\infty} df_{(+)}
\int_{f_{(+)}}^{+\infty} df_{(-)}
{\cal P}_{x}\bigl[f_{(+)} ; f_{(-)}\bigr]
\end{equation}
Let us introduce one more joint probability distribution function:
\begin{equation}
\label{17}
V_{x}(f_{1}, f_{2}) \; = \;
\mbox{Prob}\bigl[f_{(+)} > f_{1} ; \;  f_{(-)} > f_{2} \bigr] \; = \;
\int_{f_{1}}^{+\infty} df_{(+)}
\int_{f_{2}}^{+\infty} df_{(-)} \,
{\cal P}_{x}\bigl[f_{(+)} ; f_{(-)}\bigr]
\end{equation}
This two-point
free energy distribution function  gives the probability
that the free energy of the polymer with the endpoint located above the position $x$
is bigger than a given value $f_{1}$, while the free energy of the polymer with
the endpoint located below the position $x$
is bigger than a given value $f_{2}$.
According to this definition,
\begin{equation}
\label{18}
{\cal P}_{x}\bigl[f_{1} ; f_{2}\bigr] \; = \;
\frac{\partial}{\partial f_{1}}
\,
\frac{\partial}{\partial f_{2}}
\,
V_{x}(f_{1}, f_{2})
\end{equation}
Substituting this relation into eq.(\ref{16}) we find
\begin{equation}
\label{19}
W(x) \; = \;
\int_{-\infty}^{+\infty} df_{1}
\int_{f_{1}}^{+\infty} df_{2}
\frac{\partial}{\partial f_{1}}
\,
\frac{\partial}{\partial f_{2}}
\,
V_{x}(f_{1}, f_{2})
\end{equation}
Integrating by parts over $f_{2}$ and taking into account that
$V_{x}(f_{1}, f_{2})\big|_{f_{2}=+\infty} = 0$ we get
\begin{equation}
\label{20}
W(x) \; = \;
- \int_{-\infty}^{+\infty} df_{1}
\Bigl(
\frac{\partial}{\partial f_{1}}
\,
V_{x}(f_{1}, f_{2})
\Bigr)\Big|_{f_{2}=f_{1}+0}
\end{equation}
Thus, to get the distribution function $W(x)$ for the polymer's endpoint
fluctuations we have to derive the
two-point free energy distribution function $V_{x}(f_{1}, f_{2})$ first.
Note that this function is different from the
two-point free energy distribution function derived in \cite{Prolhac-Spohn}
which describes joint statistics of the free energies of the directed polymers
coming to two different endpoints.

\section{Mapping to quantum bosons}

According to the definition, eq.(\ref{17}), the probability distribution function
$V_{x}(f_{1}, f_{2})$ can be defined as follows:
\begin{equation}
\label{21}
V_{x}(f_{1}, f_{2}) = \lim_{\lambda\to\infty}
\sum_{L=0}^{\infty} \sum_{R=0}^{\infty}
\frac{(-1)^{L}}{L!} \frac{(-1)^{R}}{R!}
\exp\bigl(\lambda L f_{1} + \lambda R f_{2}\bigr) \;
\overline{\bigl[Z^{(+)}(x)\bigr]^{L} \, \bigl[Z^{(-)}(x)\bigr]^{R}}
\end{equation}
Indeed, substituting here the definitions, eqs.(\ref{13})-(\ref{14}),
we find:
\begin{eqnarray}
\nonumber
V_{x}(f_{1}, f_{2}) &=&
\lim_{\lambda\to\infty}
\int_{-\infty}^{+\infty} df_{(+)}
\int_{-\infty}^{+\infty} df_{(-)}
{\cal P}_{x}\bigl[f_{(+)} ; f_{(-)}\bigr]
\sum_{L=0}^{\infty} \sum_{R=0}^{\infty}
\frac{(-1)^{L}}{L!} \frac{(-1)^{R}}{R!} \,
\exp\bigl[\lambda L (f_{1} - f_{(+)}) \bigr] \,
\exp\bigl[\lambda R (f_{2} - f_{(-)}) \bigr]
\\
\nonumber
\\
\nonumber
&=&
\lim_{\lambda\to\infty}
\int_{-\infty}^{+\infty} df_{(+)}
\int_{-\infty}^{+\infty} df_{(-)}
{\cal P}_{x}\bigl[f_{(+)} ; f_{(-)}\bigr]
\exp\Bigl[
-\mbox{\Large e}^{\lambda (f_{1} - f_{(+)})} \,
-\mbox{\Large e}^{\lambda (f_{2} - f_{(-)})}
\Bigr]
\\
\nonumber
\\
&=&
\int_{-\infty}^{+\infty} df_{(+)}
\int_{-\infty}^{+\infty} df_{(-)}
{\cal P}_{x}\bigl[f_{(+)} ; f_{(-)}\bigr]
\theta\bigl(f_{(+)}-f_{1}\bigr) \,
\theta\bigl(f_{(-)}-f_{2}\bigr)
\label{22}
\end{eqnarray}
where $\theta(f)$ is the Heaviside step function.
We see that the above representation  coincides with the definition, eq.(\ref{17}).

Further calculations of the two-point distribution function $V_{x}(f_{1}, f_{2})$ to a large extent 
repeats the procedure described in detail in the previous paper
\cite{goe} for the one-point free energy distribution function.
Using the definitions, eqs.(\ref{13})-(\ref{14}),
the  distribution function, eq.(\ref{21}), can be represented as follows:
\begin{equation}
\label{23}
V_{x}(f_{1}, f_{2}) = \lim_{\lambda\to\infty}
\sum_{L,R=0}^{\infty}
\frac{(-1)^{L+R}}{L!R!}
\exp\bigl(\lambda L f_{1} + \lambda R f_{2}\bigr) \;
\int_{-\infty}^{x} dx_{1}...dx_{L}
\int_{x}^{+\infty} dy_{1}...dy_{R}
\Psi(x_{1},...,x_{L},y_{R},...,y_{1} ; t)
\end{equation}
where
\begin{equation}
\label{24}
\Psi(x_{1}, ..., x_{N} ; t) \; \equiv \;
\overline{Z(x_{1}) \, Z(x_{2}) \, ... \, Z(x_{N})} =
\prod_{a=1}^{N} \Biggl[\int_{\phi_a(0)=0}^{\phi_a(t)=x_a} {\cal D} \phi_a(\tau)\Biggr]
  \;  \exp\bigl(-\beta  H_{N} [\phi_{1}, \phi_{2}, ..., \phi_{N}] \bigr)
\end{equation}
with the replica Hamiltonian
\begin{equation}
\label{25}
   H_{N} [\phi_{1}, \phi_{2}, ..., \phi_{N}] \; = \;
   \frac{1}{2} \int_{0}^{t} d\tau \Biggl(
   \sum_{a=1}^{N} \bigl[\partial_\tau\phi_{a}(\tau)\bigr]^2
   - \beta u \sum_{a\not= b}^{N}
   \delta\bigl[\phi_{a}(\tau)-\phi_{b}(\tau)\bigr] \Biggr)
\end{equation}
The propagator $\Psi({\bf x}; t)$, eq.(\ref{24}), describes $N$ trajectories
$\phi_{a}(\tau)$ all starting at zero ($\phi_{a}(0) = 0$), and coming to $N$ different points
$\{x_{1}, ..., x_{N}\}$
at $\tau = t$. One can easily show that $\Psi({\bf x}; t)$
can be obtained as the solution of the  the imaginary-time
Schr\"odinger equation
\begin{equation}
   \label{26}
-\beta \, \partial_t \Psi({\bf x}; t) = \hat{H} \Psi({\bf x}; t)
\end{equation}
with the initial condition
\begin{equation}
   \label{27}
\Psi({\bf x}; 0) = \Pi_{a=1}^{N} \delta(x_a)
\end{equation}
Here the Hamiltonian is
\begin{equation}
   \label{28}
   \hat{H} =
    -\frac{1}{2}\sum_{a=1}^{N}\partial_{x_a}^2
   -\frac{1}{2}\, \kappa \sum_{a\not=b}^{N} \delta(x_a-x_b)
\end{equation}
and the interaction parameter $\kappa = \beta^{3}u$.
This Hamiltonian describes  $N$ bose-particles  interacting via
the {\it attractive} two-body potential $-\kappa \delta(x)$.

A generic  eigenstate of such system is characterized by $N$ momenta
$\{ q_{a} \} \; (a=1,...,N)$ which are splitted into
$M$  ($1 \leq M \leq N$) "clusters" described by
continuous real momenta $q_{\alpha}$ $(\alpha = 1,...,M)$
and having $n_{\alpha}$ discrete imaginary "components"
(for details see \cite{Lieb-Liniger,McGuire,Yang,Calabrese,BA-replicas,rev-TW}):
\begin{equation}
   \label{29}
q_{a} \; \equiv \; q^{\alpha}_{r} \; = \;
q_{\alpha} - \frac{i\kappa}{2}  (n_{\alpha} + 1 - 2r)
\;\; ; \; \;\;\; \;\;\; \;\;\;
(r = 1, ..., n_{\alpha})
\end{equation}
with the global constraint
\begin{equation}
   \label{30}
\sum_{\alpha=1}^{M} n_{\alpha} = N
\end{equation}
A generic solution  $\Psi({\bf x},t)$
of the Schr\"odinger equation (\ref{26}) with the initial conditions, eq.(\ref{27}),
can be represented in the form of the linear combination of the eigenfunctions
$\Psi_{\bf q}^{(M)}({\bf x})$:
\begin{equation}
   \label{31}
\Psi(x_{1}, ..., x_{N}; t) =
\sum_{M=1}^{N} \frac{1}{M!} \Biggl[\int {\cal D}^{(M)} ({\bf q},{\bf n})\Biggr] \;
|C_{M}({\bf q},{\bf n})|^{2} \;
\Psi^{(M)}_{{\bf q}}({\bf x})
{\Psi^{(M)}_{{\bf q}}}^{*}({\bf 0}) \;
\exp\bigl\{-E_{M}({\bf q},{\bf n}) t \bigr\}
\end{equation}
where we have introduced the notation
\begin{equation}
   \label{32}
\int {\cal D}^{(M)} ({\bf q},{\bf n}) \equiv
\prod_{\alpha=1}^{M} \Biggl[\int_{-\infty}^{+\infty} \frac{dq_{\alpha}}{2\pi} \sum_{n_{\alpha}=1}^{\infty}\Biggr]
{\boldsymbol \delta}\Bigl(\sum_{\alpha=1}^{M} n_{\alpha} \; , \;  N\Bigr)
\end{equation}
and ${\boldsymbol \delta}(k , m)$ is the Kronecker symbol;
note that the presence of this Kronecker symbol in the above equation
allows to extend the summations over $n_{\alpha}$'s to infinity.
Here (non-normalized) eigenfunctions are \cite{BA-replicas,rev-TW}
\begin{equation}
\label{33}
\Psi^{(M)}_{{\bf q}}({\bf x}) =
\sum_{{\cal P}}  \;
\prod_{a<b}^{N}
\Biggl[
1 +i \kappa \frac{\sgn(x_{a}-x_{b})}{q_{{\cal P}_a} - q_{{\cal P}_b}}
\Biggr] \;
\exp\Bigl[i \sum_{a=1}^{N} q_{{\cal P}_{a}} x_{a} \Bigr]
\end{equation}
where the summation goes over $N!$ permutations ${\cal P}$ of $N$ momenta $q_{a}$,
eq.(\ref{29}),  over $N$ particles $x_{a}$;
the normalization factor
\begin{equation}
   \label{34}
|C_{M}({\bf q}, {\bf n})|^{2} = \frac{\kappa^{N}}{N! \prod_{\alpha=1}^{M}\bigl(\kappa n_{\alpha}\bigr)}
\prod_{\alpha<\beta}^{M}
\frac{\big|q_{\alpha}-q_{\beta} -\frac{i\kappa}{2}(n_{\alpha}-n_{\beta})\big|^{2}}{
      \big|q_{\alpha}-q_{\beta} -\frac{i\kappa}{2}(n_{\alpha}+n_{\beta})\big|^{2}}
\end{equation}
and the eigenvalues:
\begin{eqnarray}
\nonumber
E_{M}({\bf q},{\bf n}) \; = \;
\frac{1}{2\beta} \sum_{\alpha=1}^{N} q_{a}^{2} & = &
 \frac{1}{2\beta} \sum_{\alpha=1}^{M} \; n_{\alpha} q_{\alpha}^{2}
- \frac{\kappa^{2}}{24\beta}\sum_{\alpha=1}^{M} (n_{\alpha}^{3}-n_{\alpha})
\\
\nonumber
\\
&=&
\sum_{\alpha=1}^{M} 
\Bigl[
\frac{1}{2\beta} n_{\alpha} q_{\alpha}^{2}
- \frac{\kappa^{2}}{24\beta} n_{\alpha}^{3}
\Bigr]
+ \frac{\kappa^{2}}{24\beta} N
\label{35}
\end{eqnarray}
The last term in the above expression provides just the trivial contribution to the 
selfaveraging part of the free energy (discussed in the Introduction) and therefore it will be
dropped out of the further calculations.

Using the definition, eq.(\ref{33}), one can easily prove that
\begin{equation}
\label{36}
\Psi^{(M)}_{{\bf q}}({\bf 0}) = N!
\end{equation}
In this way the problem of the calculation of the probability
distribution function, eq.(\ref{23}), reduces to the
summation over all the spectrum of the eigenstates of the $N$-particle
bosonic problem, which is parametrized by the set of both  continuous,
$\{q_{1}, ..., q_{M}\}$, and  discrete
$\{n_{1}, ...,n_{M}\}; \; (M = 1, ..., N); \; (N = 1, ..., \infty)$
degrees of freedom.

\section{Two-point free energy  distribution function}

Substituting eqs.(\ref{31})-(\ref{36}) into eq.(\ref{23}),
we get:
\begin{eqnarray}
\label{37}
V_{x}(f_{1}, f_{2}) &=& 1 + \lim_{\lambda\to\infty}
\sum_{L+R\geq 1}^{\infty} \;(-1)^{L+R} \;
\mbox{\LARGE e}^{\lambda L f_{1} + \lambda R f_{2}}
\times
\\
\nonumber
\\
\nonumber
&\times&
\sum_{M=1}^{L+R} \frac{1}{M!}
\prod_{\alpha=1}^{M}
\Biggl[
\sum_{n_{\alpha}=1}^{\infty}
\int_{-\infty}^{+\infty} \frac{dq_{\alpha}}{2\pi \kappa n_{\alpha}} \kappa^{n_{\alpha}}
\mbox{\LARGE e}^{-\frac{t}{2\beta} n_{\alpha} q_{\alpha}^{2}
+ \frac{\kappa^{2} t}{24 \beta} n_{\alpha}^{3} }
\Biggr]
\; {\boldsymbol \delta}\Bigl(\sum_{\alpha=1}^{M} n_{\alpha} \; , \;  L+R\Bigr)
\; |\tilde{C}_{M}({\bf q}, {\bf n})|^{2}
\; I_{L,R} ({\bf q}, {\bf n})
\end{eqnarray}
where
\begin{equation}
   \label{38}
|\tilde{C}_{M}({\bf q}, {\bf n})|^{2} \; = \;
\prod_{\alpha<\beta}^{M}
\frac{\big|q_{\alpha}-q_{\beta} -\frac{i\kappa}{2}(n_{\alpha}-n_{\beta})\big|^{2}}{
      \big|q_{\alpha}-q_{\beta} -\frac{i\kappa}{2}(n_{\alpha}+n_{\beta})\big|^{2}}
\end{equation}
and
\begin{eqnarray}
\nonumber
I_{L,R} ({\bf q}, {\bf n}) &=&
\sum_{{\cal P}^{(L,R)}}  \sum_{{\cal P}^{(L)}} \sum_{{\cal P}^{(R)}} \;
\prod_{a=1}^{L} \prod_{c=1}^{R}
\Biggl[
\frac{q_{{\cal P}_a^{(L)}} - q_{{\cal P}_c^{(R)}} - i \kappa}{q_{{\cal P}_a^{(L)}} - q_{{\cal P}_c^{(R)}}}
\Biggr]
\times
\prod_{a<b}^{L}\Biggl[\frac{q_{{\cal P}_a^{(L)}} - q_{{\cal P}_b^{(L)}}  - i \kappa }{
q_{{\cal P}_a^{(L)}} - q_{{\cal P}_b^{(L)}}}\Biggr]
\times
\prod_{c<d}^{R}\Biggl[\frac{q_{{\cal P}_c^{(R)}} - q_{{\cal P}_d^{(R)}}  + i \kappa }{
q_{{\cal P}_c^{(R)}} - q_{{\cal P}_d^{(R)}}}\Biggr]
\times
\\
\nonumber
\\
\nonumber
&\times&
\int_{-\infty < x_{1} \leq ... \leq x_{L}\leq x} dx_{1} ... dx_{L} \;
\exp\Bigl[i \sum_{a=1}^{L} (q_{{\cal P}_{a}^{(L)}}-i\epsilon) x_{a} \Bigr]
\\
\nonumber
\\
&\times&
\int_{x \leq y_{R} \leq ... \leq y_{1} < +\infty} dy_{R} ... dy_{1} \;
\exp\Bigl[i \sum_{c=1}^{R} (q_{{\cal P}_{c}^{(R)}}+i\epsilon) y_{c} \Bigr]
\label{39}
\end{eqnarray}
Here the summation over all permutations ${\cal P}$ of $(L+R)$ momenta
$\{q_{1}, ..., q_{L+R}\}$  over $L$ "left" particles
$\{x_{1}, ..., x_{L}\}$
and $R$ "right" particles $\{y_{R}, ..., y_{1}\}$
are divided   into three parts: the permutations ${\cal P}^{(L)}$
of $L$ momenta (taken at random out of the total list $\{q_{1}, ..., q_{L+R}\}$)
over $L$ "left" particles, the permutations ${\cal P}^{(R)}$
of the remaining $R$ momenta over $R$ "right" particles, and
finally the permutations ${\cal P}^{(L,R)}$ (or the exchange) of the
momenta between the group $"L"$ and the group $"R"$.
Note also that the integrations both over $x_{a}$'s and over $y_{c}$'s
in eq.(\ref{39}) require proper regularization at $-\infty$ and $+\infty$ correspondingly.
This is done in the standard way by introducing a supplementary parameter $\epsilon$
which will be set to zero in final results. The result of the
integrations can be represented as follows:
\begin{eqnarray}
\nonumber
I_{L,R} ({\bf q}, {\bf n}) &=& i^{-(L+R)} \;
\exp\bigl\{i x \sum_{\alpha=1}^{M} n_{\alpha} q_{\alpha} \bigr\}
\sum_{{\cal P}^{(L,R)}}  \; \;
\prod_{a=1}^{L} \prod_{c=1}^{R}
\Biggl[
\frac{q_{{\cal P}_a^{(L)}} - q_{{\cal P}_c^{(R)}} - i \kappa}{q_{{\cal P}_a^{(L)}} - q_{{\cal P}_c^{(R)}}}
\Biggr]
\times
\\
\nonumber
\\
\nonumber
&\times&
\sum_{{\cal P}^{(L)}} \; \;
\frac{1}{q^{(-)}_{{\cal P}_{1}^{(L)}} \bigl(q^{(-)}_{{\cal P}_{1}^{(L)}} + q^{(-)}_{{\cal P}_{2}^{(L)}}\bigr)... \bigl(q^{(-)}_{{\cal P}_{1}^{(L)}} + ... + q^{(-)}_{{\cal P}_{L}^{(L)}}\bigr)}
\prod_{a<b}^{L}\Biggl[\frac{q^{(-)}_{{\cal P}_a^{(L)}} - q^{(-)}_{{\cal P}_b^{(L)}}  - i \kappa }{q^{(-)}_{{\cal P}_a^{(L)}} - q^{(-)}_{{\cal P}_b^{(L)}}}\Biggr]
\times
\\
\nonumber
\\
&\times&
\sum_{{\cal P}^{(R)}} \; \;
\frac{(-1)^{R}}{q^{(+)}_{{\cal P}_{1}^{(R)}} \bigl(q^{(+)}_{{\cal P}_{1}^{(R)}} + q^{(+)}_{{\cal P}_{2}^{(R)}}\bigr)... \bigl(q^{(+)}_{{\cal P}_{1}^{(R)}} + ... + q^{(+)}_{{\cal P}_{R}^{(R)}}\bigr)}
\prod_{c<d}^{R}\Biggl[\frac{q^{(+)}_{{\cal P}_c^{(R)}} - q^{(+)}_{{\cal P}_d^{(R)}}  + i \kappa }{q^{(+)}_{{\cal P}_c^{(R)}} - q^{(+)}_{{\cal P}_d^{(R)}}}\Biggr]
\label{40}
\end{eqnarray}
where
\begin{equation}
\label{41}
q^{(\pm)}_a \; \equiv \; q_{a} \pm i\epsilon
\end{equation}
and where we have used the fact that for any permutation of the momenta, eq.(\ref{29}), one has:
\begin{equation}
 \label{42}
\sum_{a=1}^{L+R} q_{{\cal P}_{a}} \; = \;
\sum_{\alpha=1}^{M} n_{\alpha} q_{\alpha}
\end{equation}
Using the  Bethe ansatz combinatorial identity \cite{LeDoussal2},
\begin{equation}
 \label{43}
\sum_{P} \frac{1}{q_{p_{1}} (q_{p_{1}} + q_{p_{2}})... (q_{p_{1}} + ... + q _{p_{N}})}
\prod_{a<b}^{N}\Biggl[\frac{q_{p_a} - q_{p_b}  - i \kappa }{q_{p_a} - q_{p_b}}\Biggr] \; = \;
\frac{1}{\prod_{a=1}^{N} q_{a}} \;
\prod_{a<b}^{N}\Biggl[\frac{q_{a} + q_{b}  + i \kappa }{q_{a} + q_{b}}\Biggr]
\end{equation}
(where the summation goes over all permutations $P$ of $N$ momenta $\{q_{1}, ..., q_{N}\}$) we get:
\begin{eqnarray}
\nonumber
I_{L,R} ({\bf q}, {\bf n}) &=& i^{-(L+R)} \;
\exp\bigl\{i x \sum_{\alpha=1}^{M} n_{\alpha} q_{\alpha} \bigr\}
\sum_{{\cal P}^{(L,R)}}  \; \;
\prod_{a=1}^{L} \prod_{c=1}^{R}
\Biggl[
\frac{q_{{\cal P}_a^{(L)}} - q_{{\cal P}_c^{(R)}} - i \kappa}{q_{{\cal P}_a^{(L)}} - q_{{\cal P}_c^{(R)}}}
\Biggr]
\times
\\
\nonumber
\\
&\times&
\frac{1}{\prod_{a=1}^{L} q^{(-)}_{{\cal P}_{a}^{(L)}} }
\prod_{a<b}^{L}\Biggl[\frac{q^{(-)}_{{\cal P}_a^{(L)}} + q^{(-)}_{{\cal P}_b^{(L)}}  + i \kappa }{
q^{(-)}_{{\cal P}_a^{(L)}} + q^{(-)}_{{\cal P}_b^{(L)}}}\Biggr]
\times
\frac{(-1)^{R}}{\prod_{c=1}^{R} q^{(+)}_{{\cal P}_{c}^{(R)}} }
\prod_{c<d}^{R}\Biggl[\frac{q^{(+)}_{{\cal P}_c^{(R)}} + q^{(+)}_{{\cal P}_d^{(R)}}  - i \kappa }{
q^{(+)}_{{\cal P}_c^{(R)}} + q^{(+)}_{{\cal P}_d^{(R)}}}\Biggr]
\label{44}
\end{eqnarray}
Further simplification comes from the following important property of the
Bethe ansatz wave function, eq.(\ref{33}). It has such structure that
for ordered particles positions (e.g. $x_{1}<x_{2}<...<x_{N}$)
in the summation over permutations the momenta $q_{a}$ belonging
to the same cluster also remain ordered. In other words,
if we consider the momenta, eq.(\ref{29}), of a cluster $\alpha$,
$\{q_{1}^{\alpha}, q_{2}^{\alpha}, ..., q_{n_{\alpha}}^{\alpha}\}$,
belonging correspondingly to the particles $\{x_{i_{1}} < x_{i_{2}} < ... < x_{i_{n_{\alpha}}}\}$,
the permutation of any two momenta $q_{r}^{\alpha}$
and $q_{r'}^{\alpha}$ of this {\it ordered} set gives zero contribution.
Thus, in order to perform the summation over the permutations ${\cal P}^{(L,R)}$
in eq.(\ref{44}) it is sufficient to split the momenta of each cluster into two parts:
$\{q_{1}^{\alpha}, ...,  q_{m_{\alpha}}^{\alpha} ||
q_{m_{\alpha}+1}^{\alpha}..., q_{n_{\alpha}}^{\alpha}\}$, where $m_{\alpha} = 0, 1, ..., n_{\alpha}$ and
where the momenta $q_{1}^{\alpha}, ...,  q_{m_{\alpha}}^{\alpha}$ belong to the particles
of the sector $"L"$, while the momenta $q_{m_{\alpha}+1}^{\alpha}..., q_{n_{\alpha}}^{\alpha}$
belong to the particles of the sector $"R"$.

Let us introduce the numbering of the momenta
of the sector $"R"$ in the reversed order:
\begin{eqnarray}
\nonumber
q_{n_{\alpha}}^{\alpha} &\to&  {q^{*}}_{1}^{\alpha}
\\
\nonumber
q_{n_{\alpha}-1}^{\alpha} &\to&  {q^{*}}_{2}^{\alpha}
\\
\nonumber
&........&
\\
q_{m_{\alpha}+1}^{\alpha} &\to&  {q^{*}}_{s_{\alpha}}^{\alpha}
\label{45}
\end{eqnarray}
where $m_{\alpha} + s_{\alpha} = n_{\alpha}$ and (s.f. eq.(\ref{29}))
\begin{equation}
\label{46}
{q^{*}}_{r}^{\alpha} \; = \; q_{\alpha} + \frac{i \kappa}{2} (n_{\alpha} + 1 - 2r)
\; = \; q_{\alpha} + \frac{i \kappa}{2} (m_{\alpha} + s_{\alpha} + 1 - 2r)
\end{equation}
By definition, the integer parameters $\{m_{\alpha}\}$ and $\{s_{\alpha}\}$
fulfill the global constrains
\begin{eqnarray}
\label{47}
\sum_{\alpha=1}^{M} m_{\alpha} &=& L
\\
\nonumber
\\
\sum_{\alpha=1}^{M} s_{\alpha} &=& R
\label{48}
\end{eqnarray}
In this way the summation over permutations ${\cal P}^{(L,R)}$
in eq.(\ref{36}) is changed by the summations over the integer parameters
$\{m_{\alpha}\}$ and $\{s_{\alpha}\}$:
\begin{equation}
\label{49}
\sum_{{\cal P}^{(L,R)}} \; \bigl( ... \bigr) \; \to \;
\prod_{\alpha=1}^{M}
\Biggl[
\sum_{m_{\alpha}+s_{\alpha} \geq 1}^{\infty} \;
{\boldsymbol \delta}\Bigl(m_{\alpha}+s_{\alpha} \; , \;  n_{\alpha}\Bigr)
\Biggr] \;
{\boldsymbol \delta}\Bigl(\sum_{\alpha=1}^{M} m_{\alpha}\; , \; L\Bigr) \;
{\boldsymbol \delta}\Bigl(\sum_{\alpha=1}^{M} s_{\alpha}\; , \; R\Bigr)
\;
\bigl( ... \bigr)
\end{equation}
which allows to lift the summations over $L$, $R$, and $\{n_{\alpha}\}$
in eq.(\ref{37}).
Straightforward but slightly cumbersome  calculations result in the following
expression (see Appendix):
\begin{eqnarray}
 \nonumber
V_{x}(f_{1}, f_{2}) &=& \lim_{\lambda\to\infty}
\Biggl\{
1 + \sum_{M=1}^{\infty} \; \frac{(-1)^{M}}{M!} \;
\prod_{\alpha=1}^{M}
\Biggl[
\sum_{m_{\alpha}+s_{\alpha}\geq 1}^{\infty}
(-1)^{m_{\alpha}+s_{\alpha}-1}
\int_{-\infty}^{+\infty} dq_{\alpha} \;
\frac{{\cal G}\bigl(q_{\alpha}, m_{\alpha}, s_{\alpha}\bigr)}{
2\pi \kappa (m_{\alpha}+s_{\alpha})}
\times
\\
\nonumber
\\
\nonumber
&\times&
\exp\Bigl\{
-\frac{t}{2\beta} (m_{\alpha}+s_{\alpha}) q_{\alpha}^{2} +
\frac{\kappa^{2} t}{24 \beta} (m_{\alpha}+s_{\alpha})^{3} +
\lambda m_{\alpha} f_{1} + \lambda s_{\alpha} f_{2} + i x (m_{\alpha} + s_{\alpha}) q_{\alpha}
\Bigr\}
\Biggr]
\times
\\
\nonumber
\\
&\times&
|\tilde{C}_{M}({\bf q}, {\bf m + s})|^{2}  \;
{\bf G}_{M} \bigl({\bf q}, {\bf m}, {\bf s}\bigr)
\Biggr\}
\label{50}
\end{eqnarray}
where
\begin{equation}
\label{51}
|\tilde{C}_{M}({\bf q}, {\bf m + s})|^{2} \; = \;
\prod_{\alpha<\beta}^{M}
\frac{
\big|q_{\alpha}-q_{\beta}-\frac{i\kappa}{2}(m_{\alpha}+s_{\alpha}-m_{\beta}-s_{\beta})\big|^{2}}{
\big|q_{\alpha}-q_{\beta}-\frac{i\kappa}{2}(m_{\alpha}+s_{\alpha}+m_{\beta}+s_{\beta})\big|^{2}}
\end{equation}
and
\begin{equation}
\label{52}
{\cal G}\bigl(q_{\alpha}, m_{\alpha}, s_{\alpha}\bigr) \; = \;
\frac{
\Gamma\Bigl(
s_{\alpha} + \frac{2i}{\kappa} {q_{\alpha}}^{(-)}
\Bigr) \,
\Gamma\Bigl(
m_{\alpha} - \frac{2i}{\kappa} {q_{\alpha}}^{(+)}
\Bigr) \,
\Gamma\bigl(1 + m_{\alpha} + s_{\alpha}\bigr)}{
2^{(m_{\alpha} + s_{\alpha})}
\Gamma\Bigl(
m_{\alpha} + s_{\alpha} + \frac{2i}{\kappa} {q_{\alpha}}^{(-)}
\Bigr) \,
\Gamma\Bigl(
m_{\alpha} + s_{\alpha} - \frac{2i}{\kappa} {q_{\alpha}}^{(+)}
\Bigr) \,
\Gamma\bigl(1 + m_{\alpha}\bigr) \Gamma\bigl(1 + s_{\alpha}\bigr)}
\end{equation}
The explicit expression for the factor
${\bf G}_{M} \bigl({\bf q}, {\bf m}, {\bf s}\bigr)$
is given in the Appendix, eq.(\ref{A17}).

\vspace{10mm}

Redefining
\begin{equation}
\label{53}
q_{\alpha} \; = \; \frac{\kappa}{2\lambda} \, p_{\alpha}
\end{equation}
and
\begin{equation}
\label{54}
x \; \to \; \frac{2 \lambda^{2}}{\kappa} \, x
\end{equation}
with
\begin{equation}
\label{55}
\lambda \; = \; \frac{1}{2} \,
\Bigl(\frac{\kappa^{2} t}{\beta}\Bigr)^{1/3} \; = \;
\frac{1}{2} \, \bigl(\beta^{5} u^{2} t\bigr)^{1/3}
\end{equation}
the normalization factor $|\tilde{C}_{M}({\bf q}, {\bf m + s})|^{2}$, eq.(\ref{51}),
can be represented as follows:
\begin{eqnarray}
\nonumber
|\tilde{C}_{M}({\bf q}, {\bf m + s})|^{2} &=&
\prod_{\alpha<\beta}^{M}
\frac{
\big|
\lambda\bigl(m_{\alpha}+s_{\alpha}\bigr) - \lambda\bigl(m_{\beta}+s_{\beta}\bigr) -
i p_{\alpha} + ip_{\beta}
\big|^{2}}{
\big|
\lambda\bigl(m_{\alpha}+s_{\alpha}\bigr) + \lambda\bigl(m_{\beta}+s_{\beta}\bigr) -
i p_{\alpha} + ip_{\beta}
\big|^{2} }
\; = \;
\\
\nonumber
\\
\label{56}
&=&
\prod_{\alpha=1}^{M}
\bigl[2\lambda \bigl(m_{\alpha}+s_{\alpha}\bigr)\bigr]
\times
\det
\Biggl[
\frac{1}{
\lambda\bigl(m_{\alpha}+s_{\alpha}\bigr) - ip_{\alpha} +
\lambda\bigl(m_{\beta}+s_{\beta}\bigr) + ip_{\beta}}
\Biggr]_{\alpha,\beta=1,...,M}
\end{eqnarray}
where we have used the Cauchy double alternant identity
\begin{equation}
 \label{57}
\frac{\prod_{\alpha<\beta}^{M} (a_{\alpha} - a_{\beta})(b_{\alpha} - b_{\beta})}{
     \prod_{\alpha,\beta=1}^{M} (a_{\alpha} - b_{\beta})} \; = \;
(-1)^{M(M-1)/2} \det\Bigl[\frac{1}{a_{\alpha}-b_{\beta}}\Bigr]_{\alpha,\beta=1,...M}
\end{equation}
with $a_{\alpha} = p_{\alpha} - i \lambda \bigl(m_{\alpha}+s_{\alpha}\bigr)$ and
$b_{\alpha} = p_{\alpha} + i \lambda \bigl(m_{\beta}+s_{\beta}\bigr)$.

After rescaling, eqs.(\ref{53})-(\ref{55}), for the exponential factor in eq.(\ref{50}) we find
\begin{eqnarray}
\nonumber
&&
-\frac{t}{2\beta} (m_{\alpha}+s_{\alpha}) q_{\alpha}^{2}
+\frac{\kappa^{2} t}{24 \beta} (m_{\alpha}+s_{\alpha})^{3}
+ \lambda m_{\alpha} f_{1} + \lambda s_{\alpha} f_{2} + i x (m_{\alpha} + s_{\alpha}) q_{\alpha}
\; \to
\\
\nonumber
\\
&\to&
-\lambda (m_{\alpha}+s_{\alpha}) p_{\alpha}^{2}
+\frac{1}{3} \lambda^{3} (m_{\alpha}+s_{\alpha})^{3}
+\lambda m_{\alpha}f_{1} + \lambda s_{\alpha} f_{2} + i \lambda x (m_{\alpha} + s_{\alpha}) p_{\alpha}
\label{58}
\end{eqnarray}
The cubic exponential term can be linearized using the Airy function relation
\begin{equation}
   \label{59}
\exp\Bigl[ \frac{1}{3} \lambda^{3} (m_{\alpha}+s_{\alpha})^{3} \Bigr] \; = \;
\int_{-\infty}^{+\infty} dy_{\alpha} \; \Ai(y_{\alpha}) \;
\exp\Bigl[\lambda (m_{\alpha}+s_{\alpha}) \, y_{\alpha} \Bigr]
\end{equation}
Substituting eqs.(\ref{56}),(\ref{58}) and (\ref{59}) into eq.(\ref{50}), and redefining
$y_{\alpha} \; \to \; y_{\alpha} + p_{\alpha}^{2} - i x p_{\alpha}$, we get
\begin{eqnarray}
 \nonumber
V_{x}(f_{1},f_{2}) \; = \; \lim_{\lambda\to\infty}
\Biggl\{
&1& + \sum_{M=1}^{\infty} \; \frac{(-1)^{M}}{M!} \;
\prod_{\alpha=1}^{M}
\Biggl[
\int\int_{-\infty}^{+\infty} \frac{dy_{\alpha} dp_{\alpha}}{2\pi}
\Ai\bigl(y_{\alpha} + p_{\alpha}^{2} - i x p_{\alpha}\bigr)
\sum_{m_{\alpha}+s_{\alpha}\geq 1}^{\infty} (-1)^{m_{\alpha}+s_{\alpha}-1}
\times
\\
\nonumber
\\
\nonumber
&\times&
\exp\Bigl\{
\lambda m_{\alpha} (y_{\alpha} + f_{1}) +
\lambda s_{\alpha} (y_{\alpha} + f_{2})
\Bigr\} \;
{\cal G} \Bigl(\frac{p_{\alpha}}{\lambda}, \; m_{\alpha}, \; s_{\alpha}\Bigr) \;
\Biggr]
\times
\\
\nonumber
\\
&\times&
\det \hat{K}\bigl[(\lambda m_{\alpha},\, \lambda s_{\alpha}, \, p_{\alpha});
(\lambda m_{\beta}, \, \lambda s_{\beta}, \, p_{\beta})\bigr]_{\alpha,\beta=1,...,M}
\;
{\bf G}_{M} \Bigl(\frac{{\bf p}}{\lambda}, \; {\bf m}, \; {\bf s}\Bigr)
\Biggr\}
\label{60}
\end{eqnarray}
where
\begin{equation}
\label{61}
\hat{K}\bigl[(\lambda m, \, \lambda s, \, p); (\lambda m', \, \lambda s', \, p')\bigr]
\; = \;
\frac{1}{
\lambda m + \lambda s - ip +
\lambda m' + \lambda s' + ip'}
\end{equation}
The crucial point of the further calculations is the procedure of
taking the thermodynamic limit $\lambda \to \infty$. In this limit the summations over $\{m_{\alpha}\}$
and $\{s_{\alpha}\}$ are performed according to the following algorithm.
Let us consider the example of the sum of a general type:
\begin{equation}
\label{62}
R({\bf y}, {\bf p}) \; = \; \lim_{\lambda\to\infty} \prod_{\alpha=1}^{M}
\Biggl[
\sum_{n_{\alpha}=1}^{\infty} \; (-1)^{n_{\alpha} - 1}
\exp\{\lambda n_{\alpha} y_{\alpha}\}
\Biggr]\;
\Phi\Bigl(
\frac{{\bf p}}{\lambda}, \;{\bf p},\; \lambda {\bf n}, \; {\bf n}
\Bigr)
\end{equation}
where $\Phi$
is a function which depend on the factors $\lambda n_{\alpha}$, $p_{\alpha}/\lambda$
as well as on the parameters
$n_{\alpha}$ and $p_{\alpha}$ (which do not contain $\lambda$).
The summations in the above example can be represented in terms
of the integrals in the complex plane:
\begin{equation}
\label{63}
R({\bf y}, {\bf p}) \; = \; \lim_{\lambda\to\infty} \prod_{\alpha=1}^{M}
\Biggl[
\frac{1}{2i} \int_{{\cal C}} \frac{dz_{\alpha}}{\sin(\pi z_{\alpha})}
\exp\{\lambda z_{\alpha} y_{\alpha}\}
\Biggr]\;
\Phi\Bigl(
\frac{{\bf p}}{\lambda}, \;{\bf p},\; \lambda {\bf z}, \; {\bf z}
\Bigr)
\end{equation}
where the integration goes over the contour ${\cal C}$ shown in Fig.1(a).
Shifting the contour to the position ${\cal C}'$ shown in Fig.1(b)
(assuming that there is no contribution from infinity), and redefining $z \to z/\lambda$, in the
limit $\lambda \to \infty$ we get:
\begin{equation}
\label{64}
R({\bf y}, {\bf p}) \; = \;  \prod_{\alpha=1}^{M}
\Biggl[
\frac{1}{2\pi i} \int_{{\cal C}'} \frac{dz_{\alpha}}{z_{\alpha}}
\exp\{z_{\alpha} y_{\alpha}\}
\Biggr]\;
\lim_{\lambda\to\infty}
\Phi\Bigl(
\frac{{\bf p}}{\lambda}, \;{\bf p}, \; {\bf z}, \; \frac{{\bf z}}{\lambda}
\Bigr)
\end{equation}
where the parameters $y_{\alpha}$, $p_{\alpha}$ and $z_{\alpha}$
remain finite in the limit $\lambda \to \infty$.

\begin{figure}[h]
\begin{center}
   \includegraphics[width=12.0cm]{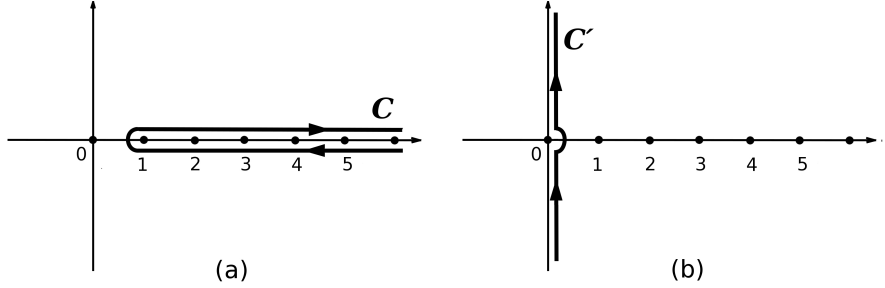}
\caption[]{The contours of integration in the complex plane used for
           summing the series:
           (a) the original contour ${\cal C}$;
           (b) the deformed contour ${\cal C}'$; }
\end{center}
\label{figure1}
\end{figure}

To perform the summations over $m_{\alpha}$ and $s_{\alpha}$ in eq.(\ref{60})
it is convenient to represent it in the following way:
\begin{equation}
 \label{65}
V_{x}(f_{1},f_{2}) \; = \;
1 + \sum_{M=1}^{\infty} \; \frac{(-1)^{M}}{M!} \;
\prod_{\alpha=1}^{M}
\Biggl[
\int\int_{-\infty}^{+\infty} \frac{dy_{\alpha} dp_{\alpha}}{2\pi}
\Ai\bigl(y_{\alpha} + p_{\alpha}^{2} - i x p_{\alpha} \bigr)
\Biggr]
\; {\cal S}_{M}\bigl({\bf p}, {\bf y}; \; f_{1}, f_{2}\bigr)
\end{equation}
where
\begin{eqnarray}
\nonumber
 {\cal S}_{M}\bigl({\bf p}, {\bf y}; \; f_{1}, f_{2}\bigr) &=&
\lim_{\lambda\to\infty}
\prod_{\alpha=1}^{M}
\Biggl[
\sum_{m_{\alpha}+s_{\alpha}\geq 1}^{\infty} (-1)^{m_{\alpha}+s_{\alpha}-1}
\exp\Bigl\{
\lambda m_{\alpha}\bigl(y_{\alpha} + f_{1}\bigr) +
\lambda s_{\alpha}\bigl(y_{\alpha} + f_{2}\bigr)
\Bigr\}
\Biggr]
\times
\\
\nonumber
\\
&\times&
\prod_{\alpha=1}^{M}
\Biggl[
{\cal G} \Bigl(\frac{p_{\alpha}}{\lambda}, \; m_{\alpha}, \; s_{\alpha}\Bigr) \;
\Biggr] \;
\det \hat{K}\bigl[(\lambda m_{\alpha},\lambda s_{\alpha},p_{\alpha});
(\lambda m_{\beta},\lambda s_{\beta}, p_{\beta})\bigr] \;
{\bf G}_{M}\Bigl(\frac{{\bf p}}{\lambda}, \; {\bf m}, \; {\bf s}\Bigr)
\label{66}
\end{eqnarray}
The summations over $m_{\alpha}$ and $s_{\alpha}$ in the above expression can
be represented as follows
\begin{equation}
\label{67}
\sum_{m_{\alpha}+s_{\alpha}\geq 1}^{\infty} (-1)^{m_{\alpha}+s_{\alpha}-1}
 =
\sum_{m_{\alpha}=1}^{\infty} (-1)^{m_{\alpha}-1} \delta(s_{\alpha}, 0) +
\sum_{s_{\alpha}=1}^{\infty} (-1)^{s_{\alpha}-1} \delta(m_{\alpha}, 0) -
\sum_{m_{\alpha}=1}^{\infty} (-1)^{m_{\alpha}-1}
\sum_{s_{\alpha}=1}^{\infty} (-1)^{s_{\alpha}-1}
\end{equation}
Thus in the integral representation, eqs.(\ref{62})-(\ref{64}),  for the function
${\cal S}_{M}\bigl({\bf p}, {\bf y}; \; f_{1}, f_{2}\bigr)$, eq.(\ref{66}), we get
\begin{eqnarray}
\nonumber
{\cal S}_{M}\bigl({\bf p}, {\bf y}; \; f_{1}, f_{2}\bigr) &=&
\prod_{\alpha=1}^{M}
\Biggl[
\int\int_{{\cal C}'}
\frac{d{z_{1}}_{\alpha}d{z_{2}}_{\alpha}}{(2\pi i)^{2}}
\Bigl(
\frac{2\pi i}{{z_{1}}_{\alpha}} \delta({z_{2}}_{\alpha}) +
\frac{2\pi i}{{z_{2}}_{\alpha}} \delta({z_{1}}_{\alpha}) -
\frac{1}{{z_{1}}_{\alpha}{z_{2}}_{\alpha}}
\Bigr)
\exp\Bigl\{
{z_{1}}_{\alpha}\bigl(y_{\alpha} + f_{1}\bigr) +
{z_{2}}_{\alpha}\bigl(y_{\alpha} + f_{2}\bigr)
\Bigr\}
\Biggr]
\times
\\
\nonumber
\\
&=&
\lim_{\lambda\to\infty}
\Biggl\{
\prod_{\alpha=1}^{M}
\Biggl[
{\cal G} \Bigl(\frac{p_{\alpha}}{\lambda}, \;\frac{{z_{1}}_{\alpha}}{\lambda} , \;\frac{{z_{2}}_{\alpha}}{\lambda} \Bigr) \;
\Biggr] \;
{\bf G}_{M}\Bigl(\frac{{\bf p}}{\lambda}, \; \frac{{\bf z_{1}}}{\lambda}, \; \frac{{\bf z_{2}}}{\lambda}\Bigr)
\Biggr\}
\;
\det \hat{K}\bigl[({z_{1}}_{\alpha},{z_{2}}_{\alpha},p_{\alpha});
({z_{1}}_{\beta},{z_{2}}_{\beta}, p_{\beta})\bigr] \;
\label{68}
\end{eqnarray}
Taking into account the Gamma function properties,
$\Gamma(z)|_{|z|\to 0} = 1/z$ and
$\Gamma(1+z)|_{|z|\to 0} = 1$, for the factors ${\cal G}$, eq.(\ref{52}), and
${\bf G}$, eq.(\ref{A17}), we obtain
\begin{equation}
\label{69}
\lim_{\lambda\to\infty}
{\cal G}\Bigl(
\frac{p_{\alpha}}{\lambda}, \;
\frac{{z_{1}}_{\alpha}}{\lambda} , \;
\frac{{z_{2}}_{\alpha}}{\lambda}
\Bigr)
\; = \;
\frac{
\bigl({z_{1}}_{\alpha}+{z_{2}}_{\alpha} + i p_{\alpha}^{(-)}\bigr)
\bigl({z_{1}}_{\alpha}+{z_{2}}_{\alpha} - i p_{\alpha}^{(+)}\bigr)}{
\bigl({z_{2}}_{\alpha} + i p_{\alpha}^{(-)}\bigr)
\bigl({z_{1}}_{\alpha} - i p_{\alpha}^{(+)}\bigr)}
\end{equation}
and
\begin{equation}
\label{70}
\lim_{\lambda\to\infty}
{\bf G}
\Bigl(
\frac{{\bf p}}{\lambda}, \; \frac{{\bf z}_{1}}{\lambda}, \;
\frac{{\bf z}_{2}}{\lambda}\Bigr)
\; = \;
1
\end{equation}
Thus, in the limit $\lambda \to \infty$ the expression for the probability distribution function, eq.(\ref{65}),
takes the form of the Fredholm determinant
\begin{eqnarray}
 \nonumber
V_{x}(f_{1},f_{2}) &=&
1 + \sum_{M=1}^{\infty} \; \frac{(-1)^{M}}{M!} \;
\prod_{\alpha=1}^{M}
\Biggl[
\int\int_{-\infty}^{+\infty} \frac{dy_{\alpha} dp_{\alpha}}{2\pi}
\Ai\bigl(y_{\alpha} + p_{\alpha}^{2} - i x p_{\alpha} \bigr)
\times
\\
\nonumber
\\
\nonumber
&\times&
\int\int_{{\cal C}'}
\frac{d{z_{1}}_{\alpha}d{z_{2}}_{\alpha}}{(2\pi i)^{2}}
\Bigl(
\frac{2\pi i}{{z_{1}}_{\alpha}} \delta({z_{2}}_{\alpha}) +
\frac{2\pi i}{{z_{2}}_{\alpha}} \delta({z_{1}}_{\alpha}) -
\frac{1}{{z_{1}}_{\alpha}{z_{2}}_{\alpha}}
\Bigr)
\Bigl(
1 + \frac{{z_{1}}_{\alpha}}{{z_{2}}_{\alpha} + i p_{\alpha}^{(-)}}
\Bigr)
\Bigl(
1 + \frac{{z_{2}}_{\alpha}}{{z_{1}}_{\alpha} - i p_{\alpha}^{(+)}}
\Bigr)
\times
\\
\nonumber
\\
\nonumber
&\times&
\exp\Bigl\{
{z_{1}}_{\alpha}\bigl(y_{\alpha} + f_{1}\bigr) +
{z_{2}}_{\alpha}\bigl(y_{\alpha} + f_{2}\bigr)
\Bigr\}
\Biggr]
\det \Bigl[
\frac{1}{{z_{1}}_{\alpha} + {z_{2}}_{\alpha} - i p_{\alpha} +
{z_{1}}_{\beta} + {z_{2}}_{\beta} + i p_{\beta}}
\Bigr]_{(\alpha,\beta)=1,2,...,M}
\\
\nonumber
\\
&=&
\det\bigl[\hat{1} \, - \, \hat{A} \bigr]
\label{71}
\end{eqnarray}
with the kernel
\begin{eqnarray}
 \nonumber
\hat{A}\bigl[({z_{1}}, \, {z_{2}}, \, p); ({z_{1}}', \, {z_{2}}', \, p')\bigr]
&=&
\int_{-\infty}^{+\infty} \frac{dy}{2\pi}
\Ai\bigl(y + p^{2} - i x p \bigr)
\Bigl(
\frac{2\pi i}{z_{1}} \delta(z_{2}) +
\frac{2\pi i}{z_{2}} \delta(z_{1}) -
\frac{1}{z_{1}{z_{2}}}
\Bigr)
\times
\\
\nonumber
\\
\nonumber
&\times&
\Bigl(
1 + \frac{z_{1}}{z_{2} + i p^{(-)}}
\Bigr)
\Bigl(
1 + \frac{z_{2}}{z_{1} - i p^{(+)}}
\Bigr)
\exp\Bigl\{
z_{1}\bigl(y + f_{1}\bigr) +
z_{2}\bigl(y + f_{2}\bigr)
\Bigr\}
\times
\\
\nonumber
\\
&\times&
\frac{1}{
{z_{1}} + {z_{2}} - ip +
{z_{1}}' + {z_{2}}' + ip'}
\label{72}
\end{eqnarray}
In the exponential representation of this determinant we get
\begin{equation}
 \label{73}
V_{x}(f_{1},f_{2}) \; = \;
\exp\Bigl[-\sum_{M=1}^{\infty} \frac{1}{M} \; \mbox{Tr} \, \hat{A}^{M} \Bigr]
\end{equation}
where
\begin{eqnarray}
 \nonumber
\mbox{Tr} \, \hat{A}^{M} &=&
\prod_{\alpha=1}^{M}
\Biggl[
\int\int_{-\infty}^{+\infty} \frac{dy_{\alpha} dp_{\alpha}}{2\pi}
\Ai\bigl(y_{\alpha} + p_{\alpha}^{2} - i x p_{\alpha} \bigr)
\times
\\
\nonumber
\\
\nonumber
&\times&
\int\int_{{\cal C}'}
\frac{d{z_{1}}_{\alpha}d{z_{2}}_{\alpha}}{(2\pi i)^{2}}
\Bigl(
\frac{2\pi i}{{z_{1}}_{\alpha}} \delta({z_{2}}_{\alpha}) +
\frac{2\pi i}{{z_{2}}_{\alpha}} \delta({z_{1}}_{\alpha}) -
\frac{1}{{z_{1}}_{\alpha}{z_{2}}_{\alpha}}
\Bigr)
\Bigl(
1 + \frac{{z_{1}}_{\alpha}}{{z_{2}}_{\alpha} + i p_{\alpha}^{(-)}}
\Bigr)
\Bigl(
1 + \frac{{z_{2}}_{\alpha}}{{z_{1}}_{\alpha} - i p_{\alpha}^{(+)}}
\Bigr)
\times
\\
\nonumber
\\
&\times&
\exp\Bigl\{
{z_{1}}_{\alpha}\bigl(y_{\alpha} + f_{1}\bigr) +
{z_{2}}_{\alpha}\bigl(y_{\alpha} + f_{2}\bigr)
\Bigr\}
\Biggr]
\;
\prod_{\alpha=1}^{M}
\Biggl[
\frac{1}{
{z_{1}}_{\alpha} + {z_{2}}_{\alpha} - i p_{\alpha} +
{z_{1}}_{\alpha +1}  + {z_{2}}_{\alpha +1}  + i p_{\alpha +1}}
\Biggr]
\label{74}
\end{eqnarray}
Here, by definition, it is assumed that ${z_{i_{M +1}}} \equiv {z_{i_{1}}}$ ($i=1,2$)
and $p_{M +1} \equiv p_{1}$.
Substituting
\begin{equation}
\label{75}
\frac{1}{
{z_{1}}_{\alpha} + {z_{2}}_{\alpha} - i p_{\alpha} +
{z_{1}}_{\alpha +1}  + {z_{2}}_{\alpha +1}  + i p_{\alpha +1}}
\; = \;
\int_{0}^{\infty} d\omega_{\alpha}
\exp\Bigl[-\bigl(
{z_{1}}_{\alpha} + {z_{2}}_{\alpha} - i p_{\alpha} +
{z_{1}}_{\alpha +1}  + {z_{2}}_{\alpha +1}  + i p_{\alpha +1}
\bigr) \, \omega_{\alpha}
\Bigr]
\end{equation}
into eq.(\ref{74}), we obtain
\begin{equation}
 \label{76}
\mbox{Tr} \, \hat{A}^{M} \; = \;
\int_{0}^{\infty} d\omega_{1} \, ... \, d\omega_{M} \,
\prod_{\alpha=1}^{M}
\Biggl[
\int\int_{-\infty}^{+\infty} \frac{dy dp}{2\pi}
\Ai\bigl(y + p^{2} + \omega_{\alpha} + \omega_{\alpha -1} - ixp \bigr)
\exp\{i p \bigl(\omega_{\alpha} - \omega_{\alpha -1}\bigr)\}
\;
S\bigl(p,  y; f_{1}, f_{2} \bigr)
\Biggr]
\end{equation}
where, by definition, $\omega_{0} \equiv \omega_{M}$, and
\begin{eqnarray}
 \nonumber
S(p, y; f_{1}, f_{2}) &=&
\int\int_{{\cal C}'}
\frac{dz_{1}dz_{2}}{(2\pi i)^{2}}
\Bigl(
\frac{2\pi i}{z_{1}} \delta(z_{2}) +
\frac{2\pi i}{z_{2}} \delta(z_{1}) -
\frac{1}{z_{1}z_{2}}
\Bigr)
\Bigl(
1 + \frac{z_{1}}{z_{2} + i p^{(-)}}
\Bigr)
\Bigl(
1 + \frac{z_{2}}{z_{1} - i p^{(+)}}
\Bigr)
\times
\\
\nonumber
\\
&\times&
\exp\Bigl\{
z_{1}\bigl(y + f_{1}\bigr) +
z_{2}\bigl(y + f_{2}\bigr)
\Bigr\}
\label{77}
\end{eqnarray}
Simple integrations  provide the following result:
\begin{eqnarray}
 \nonumber
S(p, y; f_{1}, f_{2}) &=&
  \theta(y+f_{1})
+ \theta(y+f_{2})
- \theta(y+f_{1})\theta(y+f_{2})
- \theta(y+f_{1})\theta(y+f_{2}) \exp\bigl\{i p (f_{1} - f_{2}) - 2\epsilon y\bigr\}
\\
\nonumber
\\
\nonumber
&+&
  \frac{i}{p+i\epsilon} \delta(y + f_{2})
- \frac{i}{p-i\epsilon} \delta(y + f_{1})
\\
\nonumber
\\
\nonumber
&-&
  \frac{i}{p+i\epsilon} \delta(y + f_{2}) \theta(f_{1}-f_{2})
\Bigl[1 - \exp\{i (p+i\epsilon) (f_{1} - f_{2})\}\Bigr]
\\
\nonumber
\\
&+&
  \frac{i}{p-i\epsilon} \delta(y + f_{1}) \theta(f_{2}-f_{1})
\bigl[1 - \exp\{i (p-i\epsilon) (f_{1} - f_{2})\}\bigr]
\label{78}
\end{eqnarray}
According to eq.(\ref{20}) in what follows we will be dealing with the sector $f_{2} > f_{1}$ only.
In this case the above expression  simplifies to
\begin{eqnarray}
 \nonumber
S(p, y; f_{1}, f_{2})\big|_{f_{2} > f_{1}} &=&
\Bigl(\frac{i}{p+i\epsilon} - \frac{i}{p-i\epsilon}\Bigr) \delta(y+f_{2})
\\
\nonumber
\\
\nonumber
&+&
\frac{i}{p-i\epsilon}\Bigl[\delta(y+f_{2}) - \delta(y+f_{1}) \exp\{i p (f_{1} - f_{2})\}\Bigr]
\\
\nonumber
\\
&+&
\theta(y+f_{2}) - \theta(y+f_{1}) \exp\{i p (f_{1} - f_{2})- 2\epsilon y\}
\label{79}
\end{eqnarray}
Note that at edge of the sector $f_{2} > f_{1}$ for $f_{2} = f_{1} + 0$ (in the limit $\epsilon \to 0$)
\begin{equation}
 \label{80}
S(p, y; f_{1}, f_{2})\big|_{f_{2} = f_{1}+0} \; = \; 2\pi \delta(p)\delta(y+f_{2})
\end{equation}

Thus, according to eqs.(\ref{76}) and (\ref{79}),
the two-point free energy  distribution function $V_{x}(f_{1},f_{2})$, eq.(\ref{17}),
(in the sector $f_{2} > f_{1}$) is given by the Fredholm determinant,  eq.(\ref{73}), with the kernel
\begin{eqnarray}
 \nonumber
A(\omega, \omega') &=&
\Ai(\omega + \omega' - f_{2}) -
\\
\nonumber
\\
\nonumber
&-&
\int_{-\infty}^{+\infty} \frac{dp}{2\pi}
\frac{
\Bigl[
\Ai(\omega + \omega' + p^{2} - ipx - f_{2}) -
\Ai(\omega + \omega' + p^{2} - ipx - f_{1})
\exp\{i p (f_{1} - f_{2})\}
\Bigr]}{i (p-i\epsilon)}
\exp\{i p \bigl(\omega - \omega'\bigr)\} +
\\
\nonumber
\\
\nonumber
&+&
\int_{-\infty}^{+\infty} \frac{dp}{2\pi}
\Biggl[
\int_{-f_{2}}^{+\infty} dy
\Ai(\omega + \omega' + p^{2} - ipx + y) -
\int_{-f_{1}}^{+\infty} dy
\Ai(\omega + \omega' + p^{2} - ipx + y)
\exp\{i p (f_{1} - f_{2})\}
\Biggr]
\times
\\
\nonumber
\\
&\times&
\exp\{i p \bigl(\omega - \omega'\bigr)\}
\label{81}
\end{eqnarray}
with $\omega, \omega' > 0$.

\section{The endpoint  probability distribution function}

Substituting the above result, eqs.(\ref{73}) and (\ref{81}), into eq.(\ref{20})
for the endpoint distribution function one obtains the following expression:
\begin{equation}
 \label{82}
W(x) \; = \; \int_{-\infty}^{+\infty} df \;
F_{1}(-f)
\int_{0}^{+\infty} d\omega\int_{0}^{+\infty}  d\omega'
\Bigl(\hat{1} - \hat{B}_{-f}\Bigr)^{-1} (\omega,\omega') \;
\Phi(\omega', \omega; f,x)
\end{equation}
where
\begin{equation}
 \label{83}
F_{1}(-f) = \det\bigl[\hat{1} \, - \, \hat{B}_{-f} \bigr] \; = \;
\exp\Bigl[-\sum_{M=1}^{\infty} \frac{1}{M} \; \mbox{Tr} \, \hat{B}_{-f}^{M} \Bigr]
\end{equation}
is the GOE Tracy-Widom distribution with the kernel
\begin{equation}
 \label{84}
B_{-f}(\omega, \omega') \; = \;
\Ai(\omega + \omega' - f), \; \; \; \; \; \; \; (\omega, \omega' \, > \, 0)
\end{equation}
and
\begin{eqnarray}
 \nonumber
\Phi(\omega, \omega'; f,x) &=&
i \int_{-\infty}^{+\infty} \frac{dp}{2\pi} \;
\frac{1}{p-i\epsilon} \;
\Ai'(\omega + \omega' + p^{2} - ipx - f) \;
\exp\{i p \bigl(\omega - \omega'\bigr)\}
\\
\nonumber
\\
&-&
i \int_{-f}^{+\infty} dy \; \int_{-\infty}^{+\infty} \frac{dp}{2\pi} \, p \,
\Ai(\omega + \omega' + p^{2} - ipx + y) \;
\exp\{i p \bigl(\omega - \omega'\bigr)\}
\label{85}
\end{eqnarray}
Using the standard integral representation of the Airy function one can easily reduce the above function
$\Phi(\omega, \omega'; f,x)$ to the following sufficiently simple form:
\begin{eqnarray}
 \nonumber
\Phi(\omega, \omega'; f,x) = -\frac{1}{2} \int_{0}^{+\infty} dy \;
&\Biggl[&
\Bigl(
\frac{\partial}{\partial \omega} - \frac{\partial}{\partial \omega'}
\Bigr)
\Psi\bigl(\omega - \frac{1}{2}f + y ; \; x \bigr)
\Psi\bigl(\omega' - \frac{1}{2}f + y ; \; -x\bigr) +
\\
\nonumber
\\
&+&
\Bigl(
\frac{\partial}{\partial \omega} + \frac{\partial}{\partial \omega'}
\Bigr)
\Psi\bigl(\omega - \frac{1}{2}f - y ; \; x\bigr)
\Psi\bigl(\omega' - \frac{1}{2}f + y ; \; -x\bigr)
\Biggr]
\label{86}
\end{eqnarray}
where 
\begin{equation}
 \label{87}
\Psi(\omega; x) \; = \;
2^{1/3} \mbox{Ai}\Bigl[2^{1/3}\bigl(\omega + \frac{1}{8} x^{2}\bigr)\Bigr] \, 
\exp\bigl\{-\frac{1}{2} \omega x\bigr\}
\end{equation}

Thus, eqs.(\ref{82}), (\ref{86}) and (\ref{87}) complete the derivation of the
probability distribution function for the directed polymer's endpoint.
Unfortunately, at present stage the analytical properties of this function are
not quite clear. The study of this function require the special analysis and it
will be done elsewhere.

\section{Conclusions}

In this paper the explicit expression for the the probability
distribution function of the endpoint of one-dimensional directed polymers
in random potential is derived  in terms of the Bethe ansatz replica
technique. The result obtained, eqs.(\ref{82})-(\ref{87}),
looks quite similar to the one derived
in terms of completely different method in which the
maximal point of the $\mbox{Airy}_{2}$ process minus a parabola
is considered \cite{math1, math2, math3}.
Unfortunately, at present stage the final expression for the
probability distribution function obtained both here
and in Refs.\cite{math1, math2, math3} is rather sophisticated
so that the study of its analytical properties would require
special efforts. Hopefully this problem will be solved in the near future.

One more conclusion of the present study is that the approach used,
namely the Bethe ansatz replica technique, once again
(following the works \cite{BA-replicas, LeDoussal1, LeDoussal2, goe})
has demonstrated its efficiency. Hopefully it will also be
fruitful  for the studies of more serious problems in this scope,
such as joint statistical properties of the free energy fluctuations at different times.

\acknowledgments

This work was supported in part by the grant IRSES DCPA PhysBio-269139.

\vspace{15mm}

\begin{center}

\appendix{\Large Appendix }

\end{center}

\newcounter{A}
\setcounter{equation}{0}
\renewcommand{\theequation}{A.\arabic{equation}}

\vspace{10mm}

 In terms of the parameters $\{m_{\alpha}\}$ and $\{s_{\alpha}\}$
the product factors in eq.(\ref{44}) are expressed as follows:
\begin{eqnarray}
\label{A1}
\prod_{a=1}^{L} q^{(-)}_{{\cal P}_{a}^{(L)}}
&=&
\prod_{\alpha=1}^{M} \prod_{r=1}^{m_{\alpha}}
{q^{\alpha}_{r}}^{(-)}
\\
\nonumber
\\
\label{A2}
\prod_{a=1}^{R} q^{(+)}_{{\cal P}_{a}^{(R)}}
&=&
\prod_{\alpha=1}^{M} \prod_{r=1}^{s_{\alpha}}
{{q^{*}}^{\alpha}_{r}}^{(+)}
\end{eqnarray}
\begin{eqnarray}
\label{A3}
\prod_{a<b}^{L}
\Biggl[
\frac{
q^{(-)}_{{\cal P}_a^{(L)}} + q^{(-)}_{{\cal P}_b^{(L)}}  + i \kappa }{
q^{(-)}_{{\cal P}_a^{(L)}} + q^{(-)}_{{\cal P}_b^{(L)}}}
\Biggr]
&=&
\prod_{\alpha=1}^{M} \prod_{1\leq r< r'}^{m_{\alpha}}
\Biggl[
\frac{
{q^{\alpha}_{r}}^{(-)}+{q^{\alpha}_{r'}}^{(-)}+i\kappa}{
{q^{\alpha}_{r}}^{(-)}+{q^{\alpha}_{r'}}^{(-)}}
\Biggr]
\times
\prod_{1\leq\alpha<\beta}^{M} \prod_{r=1}^{m_{\alpha}}\prod_{r'=1}^{m_{\beta}}
\Biggl[
\frac{
{q^{\alpha}_{r}}^{(-)}+{q^{\beta}_{r'}}^{(-)}+i\kappa}{
{q^{\alpha}_{r}}^{(-)}+{q^{\beta}_{r'}}^{(-)}}
\Biggr]
\\
\nonumber
\\
\nonumber
\\
\label{A4}
\prod_{c<d}^{R}
\Biggl[
\frac{
q^{(+)}_{{\cal P}_c^{(R)}} + q^{(+)}_{{\cal P}_d^{(R)}}-i \kappa }{
q^{(+)}_{{\cal P}_c^{(R)}} + q^{(+)}_{{\cal P}_d^{(R)}}}\Biggr]
&=&
\prod_{\alpha=1}^{M} \prod_{1\leq r< r'}^{s_{\alpha}}
\Biggl[
\frac{
{{q^{*}}^{\alpha}_{r}}^{(+)}+{{q^{*}}^{\alpha}_{r'}}^{(+)}-i\kappa}{
{{q^{*}}^{\alpha}_{r}}^{(+)}+{{q^{*}}^{\alpha}_{r'}}^{(+)}}
\Biggr]
\times
\prod_{1\leq\alpha<\beta}^{M} \prod_{r=1}^{s_{\alpha}}\prod_{r'=1}^{s_{\beta}}
\Biggl[
\frac{
{{q^{*}}^{\alpha}_{r}}^{(+)}+{{q^{*}}^{\beta}_{r'}}^{(+)}-i\kappa}{
{{q^{*}}^{\alpha}_{r}}^{(+)}+{{q^{*}}^{\beta}_{r'}}^{(+)}}
\Biggr]
\\
\nonumber
\\
\nonumber
\\
\nonumber
\prod_{a=1}^{L} \prod_{c=1}^{R}
\Biggl[
\frac{
q_{{\cal P}_a^{(L)}} - q_{{\cal P}_c^{(R)}} - i\kappa}{
q_{{\cal P}_a^{(L)}} - q_{{\cal P}_c^{(R)}}}
\Biggr]
&=&
\prod_{1\leq\alpha<\beta}^{M}
\Biggl\{
\prod_{r=1}^{m_{\alpha}}\prod_{r'=1}^{s_{\beta}}
\Biggl[
\frac{
q^{\alpha}_{r} - {q^{*}}^{\beta}_{r'} - i\kappa}{
q^{\alpha}_{r} + {q^{*}}^{\beta}_{r'}  }
\Biggr]
\times
\prod_{r=1}^{s_{\alpha}}\prod_{r'=1}^{m_{\beta}}
\Biggl[
\frac{
{q^{*}}^{\alpha}_{r} - q^{\beta}_{r'} - i\kappa}{
{q^{*}}^{\alpha}_{r} - q^{\beta}_{r'}  }
\Biggr]
\Biggr\}
\times
\\
\nonumber
\\
\nonumber
\\
\label{A5}
&\times&
\prod_{\alpha=1}^{M} \prod_{r=1}^{m_{\alpha}}\prod_{r'=1}^{s_{\alpha}}
\Biggl[
\frac{
q^{\alpha}_{r} - {q^{*}}^{\alpha}_{r'} - i\kappa}{
q^{\alpha}_{r} - {q^{*}}^{\alpha}_{r'}  }
\Biggr]
\end{eqnarray}
Substituting eqs.(\ref{A1})-(\ref{A5}) into eq.(\ref{44}),
and then substituting the resulting expression into eq.(\ref{37})
we obtain eq.(\ref{50}) where
\begin{eqnarray}
\nonumber
{\cal G}\bigl(q_{\alpha}, m_{\alpha}, s_{\alpha}\bigr) &=&
\frac{(-1)^{s_{\alpha}} (-i\kappa)^{(m_{\alpha}+s_{\alpha})}}{
\prod_{r=1}^{m_{\alpha}}{q^{\alpha}_{r}}^{(-)}
\prod_{r=1}^{s_{\alpha}}{{q^{*}}^{\alpha}_{r}}^{(+)}}
\times
\\
\nonumber
\\
&\times&
\prod_{r< r'}^{m_{\alpha}}
\Biggl[
\frac{
{q^{\alpha}_{r}}^{(-)}+{q^{\alpha}_{r'}}^{(-)}+i\kappa}{
{q^{\alpha}_{r}}^{(-)}+{q^{\alpha}_{r'}}^{(-)}}
\Biggr]
\prod_{r< r'}^{s_{\alpha}}
\Biggl[
\frac{
{{q^{*}}^{\alpha}_{r}}^{(+)}+{{q^{*}}^{\alpha}_{r'}}^{(+)}-i\kappa}{
{{q^{*}}^{\alpha}_{r}}^{(+)}+{{q^{*}}^{\alpha}_{r'}}^{(+)}}
\Biggr]
\prod_{r=1}^{m_{\alpha}}\prod_{r'=1}^{s_{\alpha}}
\Biggl[
\frac{
q^{\alpha}_{r} - {q^{*}}^{\alpha}_{r'} - i\kappa}{
q^{\alpha}_{r} - {q^{*}}^{\alpha}_{r'}  }
\Biggr]
\label{A6}
\end{eqnarray}
and
\begin{eqnarray}
\nonumber
{\bf G}_{M} \bigl({\bf q}, {\bf m}, {\bf s}\bigr) &=& 
\prod_{1\leq\alpha<\beta}^{M}
\Biggl\{
\prod_{r=1}^{m_{\alpha}}\prod_{r'=1}^{m_{\beta}}
\Biggl[
\frac{
{q^{\alpha}_{r}}^{(-)}+{q^{\beta}_{r'}}^{(-)}+i\kappa}{
{q^{\alpha}_{r}}^{(-)}+{q^{\beta}_{r'}}^{(-)}}
\Biggr]
\prod_{r=1}^{s_{\alpha}}\prod_{r'=1}^{s_{\beta}}
\Biggl[
\frac{
{{q^{*}}^{\alpha}_{r}}^{(+)}+{{q^{*}}^{\beta}_{r'}}^{(+)}-i\kappa}{
{{q^{*}}^{\alpha}_{r}}^{(+)}+{{q^{*}}^{\beta}_{r'}}^{(+)}}
\Biggr]
\times
\\
\nonumber
\\
&\times&
\prod_{r=1}^{m_{\alpha}}\prod_{r'=1}^{s_{\beta}}
\Biggl[
\frac{
q^{\alpha}_{r} - {q^{*}}^{\beta}_{r'} - i\kappa}{
q^{\alpha}_{r} + {q^{*}}^{\beta}_{r'}  }
\Biggr]
\times
\prod_{r=1}^{s_{\alpha}}\prod_{r'=1}^{m_{\beta}}
\Biggl[
\frac{
{q^{*}}^{\alpha}_{r} - q^{\beta}_{r'} - i\kappa}{
{q^{*}}^{\alpha}_{r} - q^{\beta}_{r'}  }
\Biggr]
\Biggr\}
\label{A7}
\end{eqnarray}
The product factors in eq.(\ref{A6}) can be easily expressed it terms of the
Gamma functions:
\begin{eqnarray}
\label{A8}
\prod_{r=1}^{m_{\alpha}}{q^{\alpha}_{r}}^{(-)} &=&
\prod_{r=1}^{m_{\alpha}}
\Bigl[
{q_{\alpha}}^{(-)} - \frac{i\kappa}{2}(m_{\alpha}+s_{\alpha}+1) + i\kappa r
\Bigr]
\; = \;
(i\kappa)^{m_{\alpha}}
\frac{
\Gamma\Bigl(
\frac{1}{2}-\frac{s_{\alpha}-m_{\alpha}}{2} - \frac{i{q_{\alpha}}^{(-)}}{\kappa}
\Bigr)}{
\Gamma\Bigl(
\frac{1}{2}-\frac{s_{\alpha}+m_{\alpha}}{2} - \frac{i{q_{\alpha}}^{(-)}}{\kappa}
\Bigr)}
\\
\nonumber
\\
\nonumber
\\
\prod_{r=1}^{s_{\alpha}}{{q^{*}}^{\alpha}_{r}}^{(+)} &=&
\prod_{r=1}^{s_{\alpha}}
\Bigl[
{q_{\alpha}}^{(+)} + \frac{i\kappa}{2}(m_{\alpha}+s_{\alpha}+1) - i\kappa r
\Bigr]
\; = \;
(-i\kappa)^{s_{\alpha}}
\frac{
\Gamma\Bigl(
\frac{1}{2}-\frac{m_{\alpha}-s_{\alpha}}{2} + \frac{i{q_{\alpha}}^{(+)}}{\kappa}
\Bigr)}{
\Gamma\Bigl(
\frac{1}{2}-\frac{m_{\alpha}+s_{\alpha}}{2} + \frac{i{q_{\alpha}}^{(+)}}{\kappa}
\Bigr)}
\label{A9}
\end{eqnarray}
\begin{eqnarray}
\label{A10}
\prod_{r< r'}^{m_{\alpha}}
\Biggl[
\frac{
{q^{\alpha}_{r}}^{(-)}+{q^{\alpha}_{r'}}^{(-)}+i\kappa}{
{q^{\alpha}_{r}}^{(-)}+{q^{\alpha}_{r'}}^{(-)}}
\Biggr]
&=&
2^{-(m_{\alpha}-1)}
\frac{
\Gamma\Bigl(
m_{\alpha}-s_{\alpha} - \frac{2i{q_{\alpha}}^{(-)}}{\kappa}
\Bigr)
\Gamma\Bigl(
1-\frac{m_{\alpha}+s_{\alpha}}{2} - \frac{i{q_{\alpha}}^{(-)}}{\kappa}
\Bigr)}{
\Gamma\Bigl(
\frac{m_{\alpha}-s_{\alpha}}{2} - \frac{i{q_{\alpha}}^{(-)}}{\kappa}
\Bigr)
\Gamma\Bigl(
1 - s_{\alpha} - \frac{2i{q_{\alpha}}^{(-)}}{\kappa}
\Bigr)}
\\
\nonumber
\\
\nonumber
\\
\label{A11}
\prod_{r< r'}^{s_{\alpha}}
\Biggl[
\frac{
{{q^{*}}^{\alpha}_{r}}^{(+)}+{{q^{*}}^{\alpha}_{r'}}^{(+)}-i\kappa}{
{{q^{*}}^{\alpha}_{r}}^{(+)}+{{q^{*}}^{\alpha}_{r'}}^{(+)}}
\Biggr]
&=&
2^{-(s_{\alpha}-1)}
\frac{
\Gamma\Bigl(
s_{\alpha}-m_{\alpha} + \frac{2i{q_{\alpha}}^{(+)}}{\kappa}
\Bigr)
\Gamma\Bigl(
1-\frac{m_{\alpha}+s_{\alpha}}{2} + \frac{i{q_{\alpha}}^{(+)}}{\kappa}
\Bigr)}{
\Gamma\Bigl(
\frac{s_{\alpha}-m_{\alpha}}{2} + \frac{i{q_{\alpha}}^{(+)}}{\kappa}
\Bigr)
\Gamma\Bigl(
1 - m_{\alpha} + \frac{2i{q_{\alpha}}^{(+)}}{\kappa}
\Bigr)}
\\
\nonumber
\\
\nonumber
\\
\label{A12}
\prod_{r=1}^{m_{\alpha}}\prod_{r'=1}^{s_{\alpha}}
\Biggl[
\frac{
q^{\alpha}_{r} - {q^{*}}^{\alpha}_{r'} - i\kappa}{
q^{\alpha}_{r} - {q^{*}}^{\alpha}_{r'}  }
\Biggr]
&=&
\frac{
\Gamma\bigl(1 + m_{\alpha} + s_{\alpha}\bigr)}{
\Gamma\bigl(1 + m_{\alpha}\bigr) \Gamma\bigl(1 + s_{\alpha}\bigr)}
\end{eqnarray}
Substituting the above expressions into eq.(\ref{A6}) and using the
standard relations for the Gamma functions,
\begin{eqnarray}
\label{A13}
\Gamma(z) \,\Gamma(1-z) &=& \frac{\pi}{\sin(\pi z)}
\\
\nonumber
\\
\label{A14}
\Gamma(1+z) &=& z \, \Gamma(z)
\\
\nonumber
\\
\label{A15}
\Gamma\Bigl(\frac{1}{2} + z\Bigr) &=&
\frac{\sqrt{\pi} \, \Gamma\bigl(1 + 2z\bigr)}{
2^{2z} \, \Gamma\bigl(1 + z\bigr)}
\end{eqnarray}
for the factor ${\cal G}$, eq.(\ref{A6}), we get
\begin{equation}
\label{A16}
{\cal G}\bigl(q_{\alpha}, m_{\alpha}, s_{\alpha}\bigr) \; = \;
\frac{
\Gamma\Bigl(
s_{\alpha} + \frac{2i}{\kappa} {q_{\alpha}}^{(-)}
\Bigr) \,
\Gamma\Bigl(
m_{\alpha} - \frac{2i}{\kappa} {q_{\alpha}}^{(+)}
\Bigr) \,
\Gamma\bigl(1 + m_{\alpha} + s_{\alpha}\bigr)}{
2^{(m_{\alpha} + s_{\alpha})}
\Gamma\Bigl(
m_{\alpha} + s_{\alpha} + \frac{2i}{\kappa} {q_{\alpha}}^{(-)}
\Bigr) \,
\Gamma\Bigl(
m_{\alpha} + s_{\alpha} - \frac{2i}{\kappa} {q_{\alpha}}^{(+)}
\Bigr) \,
\Gamma\bigl(1 + m_{\alpha}\bigr) \Gamma\bigl(1 + s_{\alpha}\bigr)}
\end{equation}
Similar calculations for the factor ${\bf G}_{M}$, eq.(\ref{A7})
yield the following expression
\begin{eqnarray}
\nonumber
{\bf G}_{M} \bigl({\bf q}, {\bf m}, {\bf s}\bigr) &=&
\prod_{1\leq\alpha<\beta}^{M}
\Biggl\{
\frac{
\Gamma
\Bigl[
1 + \frac{m_{\alpha} + m_{\beta} - s_{\alpha} - s_{\beta}}{2}
-\frac{i}{\kappa}\bigl({q_{\alpha}}^{(-)} + {q_{\beta}}^{(-)}\bigr)
\Bigr] \,
\Gamma
\Bigl[
1 - \frac{m_{\alpha} + m_{\beta} + s_{\alpha} + s_{\beta}}{2}
-\frac{i}{\kappa}\bigl({q_{\alpha}}^{(-)} + {q_{\beta}}^{(-)}\bigr)
\Bigr]}{
\Gamma
\Bigl[
1 - \frac{m_{\alpha} - m_{\beta} + s_{\alpha} + s_{\beta}}{2}
-\frac{i}{\kappa}\bigl({q_{\alpha}}^{(-)} + {q_{\beta}}^{(-)}\bigr)
\Bigr] \,
\Gamma
\Bigl[
1 + \frac{m_{\alpha} - m_{\beta} - s_{\alpha} - s_{\beta}}{2}
-\frac{i}{\kappa}\bigl({q_{\alpha}}^{(-)} + {q_{\beta}}^{(-)}\bigr)
\Bigr]}
\times
\\
\nonumber
\\
\nonumber
\\
\nonumber
&\times&
\frac{
\Gamma
\Bigl[
1 - \frac{m_{\alpha} + m_{\beta} - s_{\alpha} - s_{\beta}}{2}
+\frac{i}{\kappa}\bigl({q_{\alpha}}^{(+)} + {q_{\beta}}^{(+)}\bigr)
\Bigr] \,
\Gamma
\Bigl[
1 - \frac{m_{\alpha} + m_{\beta} + s_{\alpha} + s_{\beta}}{2}
+\frac{i}{\kappa}\bigl({q_{\alpha}}^{(+)} + {q_{\beta}}^{(+)}\bigr)
\Bigr]}{
\Gamma
\Bigl[
1 - \frac{m_{\alpha} + m_{\beta} + s_{\alpha} - s_{\beta}}{2}
+\frac{i}{\kappa}\bigl({q_{\alpha}}^{(+)} + {q_{\beta}}^{(+)}\bigr)
\Bigr] \,
\Gamma
\Bigl[
1 - \frac{m_{\alpha} + m_{\beta} - s_{\alpha} + s_{\beta}}{2}
+\frac{i}{\kappa}\bigl({q_{\alpha}}^{(+)} + {q_{\beta}}^{(+)}\bigr)
\Bigr]}
\times
\\
\nonumber
\\
\nonumber
\\
\nonumber
&\times&
\frac{
\Gamma
\Bigl[
1 + \frac{m_{\alpha} + m_{\beta} + s_{\alpha} + s_{\beta}}{2}
+\frac{i}{\kappa}\bigl(q_{\alpha} - q_{\beta}\bigr)
\Bigr] \,
\Gamma
\Bigl[
1 + \frac{-m_{\alpha} + m_{\beta} + s_{\alpha} - s_{\beta}}{2}
+\frac{i}{\kappa}\bigl(q_{\alpha} - q_{\beta}\bigr)
\Bigr]}{
\Gamma
\Bigl[
1 + \frac{-m_{\alpha} + m_{\beta} + s_{\alpha} + s_{\beta}}{2}
+\frac{i}{\kappa}\bigl(q_{\alpha} - q_{\beta}\bigr)
\Bigr] \,
\Gamma
\Bigl[
1 + \frac{m_{\alpha} + m_{\beta} + s_{\alpha} - s_{\beta}}{2}
+\frac{i}{\kappa}\bigl(q_{\alpha} - q_{\beta}\bigr)
\Bigr]}
\times
\\
\nonumber
\\
\nonumber
\\
\label{A17}
&\times&
\frac{
\Gamma
\Bigl[
1 + \frac{m_{\alpha} + m_{\beta} + s_{\alpha} + s_{\beta}}{2}
-\frac{i}{\kappa}\bigl(q_{\alpha} - q_{\beta}\bigr)
\Bigr] \,
\Gamma
\Bigl[
1 + \frac{m_{\alpha} - m_{\beta} - s_{\alpha} + s_{\beta}}{2}
-\frac{i}{\kappa}\bigl(q_{\alpha} - q_{\beta}\bigr)
\Bigr]}{
\Gamma
\Bigl[
1 + \frac{m_{\alpha} + m_{\beta} - s_{\alpha} + s_{\beta}}{2}
-\frac{i}{\kappa}\bigl(q_{\alpha} - q_{\beta}\bigr)
\Bigr] \,
\Gamma
\Bigl[
1 + \frac{m_{\alpha} - m_{\beta} + s_{\alpha} + s_{\beta}}{2}
-\frac{i}{\kappa}\bigl(q_{\alpha} - q_{\beta}\bigr)
\Bigr]}
\Biggr\}
\end{eqnarray}

\newpage


\end{document}